\documentclass[aps,prb,twocolumn,showpacs,superscriptaddress,longbibliography]{revtex4-2}

\usepackage{amssymb}
\usepackage{amsfonts}
\usepackage{amsmath}
\usepackage{subfigure}
\usepackage{multirow}
\usepackage{bm}
\usepackage{textcomp}
\usepackage{color}
\usepackage{graphicx}
\usepackage{float}
\usepackage{parskip}
\usepackage{textcomp}

\usepackage[a4paper=true,pagebackref=false]{hyperref}
\hypersetup{colorlinks,citecolor=blue,filecolor=blue,linkcolor=blue,urlcolor=blue}

\usepackage[normalem]{ulem}
\usepackage{hyperref}
\usepackage[english]{babel}

\begin{document}

\title{Introducing the step Monte Carlo method for simulating dynamic properties}

\author{D.~Sztenkiel} \email{sztenkiel@ifpan.edu.pl}
\affiliation{Institute of Physics, Polish Academy of Sciences, Aleja Lotnikow 32/46, PL-02668 Warsaw, Poland}

\date{\today}

\begin{abstract}


In this work, we introduce a simple modification of the Monte Carlo algorithm, which we call step Monte Carlo (sMC). The sMC approach allows to simulate processes far from equilibrium and obtain information about the dynamic properties of the system under investigation. In the approach proposed here the probability of accepting the final (trial) state depends on the activation energy, not on the relative energy between the final and initial state. This barrier height is probed on an ongoing basis, by generating intermediate states along the path connecting the initial and trial positions. Importantly, to calculate the activation energy, our model only requires knowledge of the Hamiltonian without having to introduce additional input parameters such as transition rates etc. The details of sMC are explained for the case of a simple spin model. The comparison of its results with the ones obtained within the frame of stochastic Landau-Lifshitz-Gilbert indicates the correctness of sMC. In our opinion, the proposed here method can be applied to simulate other processes, for example dynamics of classical atoms and complex fluids, diffusion, nucleation, surface adsorption and crystal growth processes.

	
\end{abstract}

\maketitle

\section{Introduction}

\baselineskip 15pt

The Monte Carlo (MC) method has proved to be a valuable simulation tool in many branches of science such as physics, chemistry \cite{Liu:2004_Springer}, biology \cite{Liu:2004_Springer}, computer science \cite{Liu:2004_Springer}, economics - finance \cite{Glasserman:2003_Springer, Gerstner:2013_SingaporeWS} and engineering \cite{Fishman:2003_Springer}. In the field of condensed matter physics and materials science MC can be used to study, among many others, system of classical particles \cite{Frenkel:2002_book, Leach:1996_book},  classical spin systems \cite{Sawicki:2012_PRB, Simserides:2014_EPJ, Evans:2014_JPhysCM, Sato:2018_PRB}, nucleation and crystal growth processes \cite{Sitter:1995_TSF, vanHoof:1998_JCG, Woodraska:1997_SurfScience, ZaluskaKotur:2013_CGD, Song:2021_NL}, polymer solutions \cite{Urakami:1996_JPSJ}, hopping transport \cite{Murayama:1997_SSC,Ma:2007_PhysB, Dimakogianni:2013_PM}, percolation \cite{Dhar:1981_JPCSSP} and fractals problems. The MC algorithm is the natural choice for studying the static properties of a system, where dynamical effects are not required. Then, the advantages of Monte Carlo are the relative ease of implementation and the rapid convergence to steady state. In general, the theory of equilibrium properties is well developed for a wide variety of models and materials. Here, however, we shall consider situations far from equilibrium. For example, the MBE growth of thin films proceeds by both deposition and diffusion processes. The probability of an atom to hop to a nearest empty site, does not depend only on the relative energies of the configuration before and after the hopping event, but rather on the barrier height. The Kinetic Monte Carlo (KMC) method \cite{Young:1966_PPS, Bortz:1975_JCP} was developed for evolving systems dynamically from state to state. KMC simulates how the occupation of the sites changes over time in a system with a known transition rates (these rates are inputs to the KMC algorithm). 

In the modified MC approach proposed here (similarly to the KMC approach) the probability of accepting the final state depends on the activation energy, not on the relative energy between the final and initial state. However, the barrier height is calculated on an ongoing basis, by generating intermediate states with a predefined step $\Delta$. Therefore, we name this method step Monte Carlo (sMC). Importantly, the sMC method correctly takes into account the presence of various local barriers and it obeys the detailed balance condition, even if the system is not in equilibrium. As a result, the appropriate dynamics of the tested system is simulated. The details of sMC algorithm are explained for the case of magnetization process. To test the correctness of sMC, we compare its results with those obtained by stochastic Landau-Lifshitz-Gilbert (sLLG) approach. 

\section{Step Monte-Carlo method}

\begin{figure*}[htp]
    \centering
		\includegraphics[width=12cm]{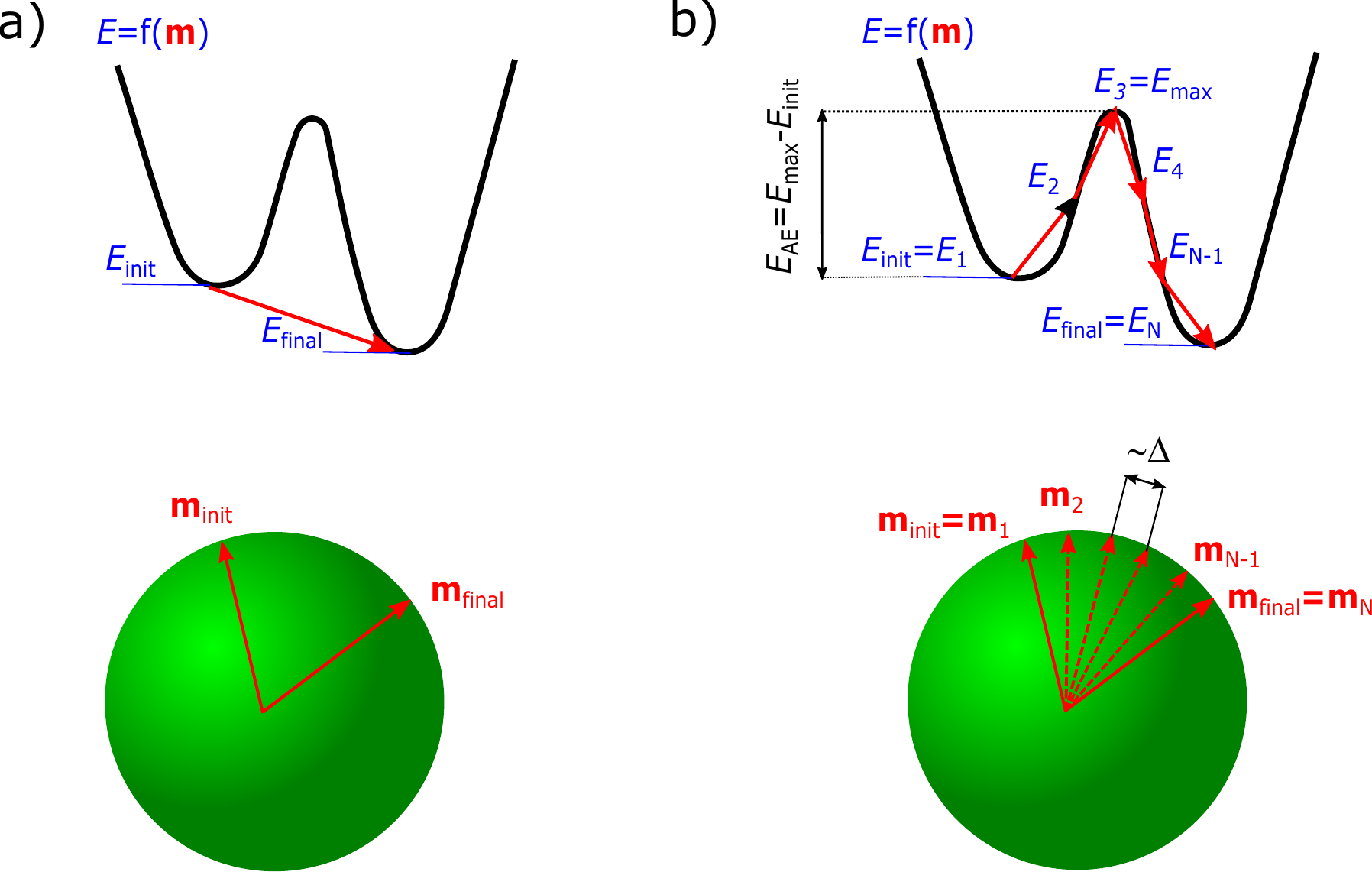}
    \caption{Schematic illustration of (a) Monte Carlo and (b) step Monte Carlo (sMC) methods for the case of magnetization process (one Monte Carlo move). The magnetization direction $\textbf{m}$ (bottom panels) can take any point on the unit sphere with energy $E=f(\textbf{m})$ (top panels). $\textbf{m}_{init}$ and $\textbf{m}_{final}$ are initial and final (trial) direction of magnetic moment with energy E$_{init}$ and E$_{final}$ respectively. In the sMC approach the probability of accepting the final  random state $\textbf{m}_{final}$ depends on the activation energy  $E_{AE}$ = $E_{\mathrm{max}}$ - $E_{init}$, not on the relative energy $E_{final}$ - $E_{init}$. This barrier height is probed by generating intermediate magnetic states $\textbf{m}_k$ ($k=2, 3, ..., N-1$) on the arc connecting the initial $\textbf{m}_{init}=\textbf{m}_1$ and the final $\textbf{m}_{final}=\textbf{m}_N$ magnetization direction with a predefined magnetization step $\Delta$. This means that $\textbf{m}_{k+1}$ is obtained in such a way that $|\textbf{m}_{k+1}-\textbf{m}_{k}| \propto \Delta$.}
    \label{fig:sMC_schematic}
\end{figure*}

\begin{figure}[htp]
    \centering
		\includegraphics[width=6.0cm]{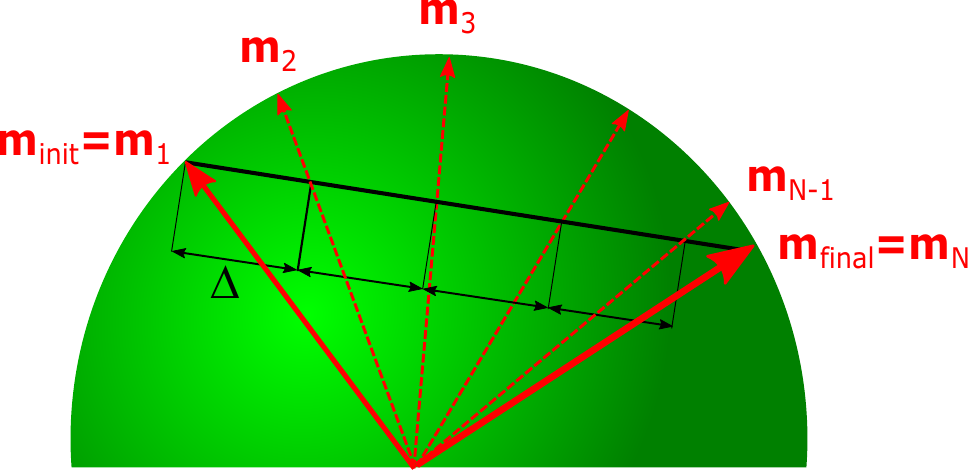}
    \caption{Schematic illustration of the generation of intermediate states by shifting the magnetization direction on the unit sphere (on the arc connecting the initial $\textbf{m}_{init}$ and the final $\textbf{m}_{final}$ magnetization direction).}
    \label{fig:IntermediateStates}
\end{figure}


Here we outline the details of step Monte Carlo method, that allows us to simulate processes far from equilibrium and obtain information about the dynamic properties of the system under investigation. The differences between sMC and standard MC are shown schematically in Fig.~\ref{fig:sMC_schematic} for the case of the magnetization process. In sMC, the probability of accepting the final magnetic direction $\textbf{m}_{final}$ depends on the activation energy (barrier height) $E_{AE}$, not on the relative energy between the final and initial state $E_{final}$ - $E_{inital}$. To calculate the barrier height, we probe energies $E_k=f(\textbf{m}_k)$ of intermediate magnetic states $\textbf{m}_2$, $\textbf{m}_3$,...,$\textbf{m}_{N-1}$. These intermediate states are generated on the arc connecting the initial $\textbf{m}_{init}=\textbf{m}_1$ and the final (trial) $\textbf{m}_{final}=\textbf{m}_N$ magnetization direction. This means that $\textbf{m}_{k+1}$ is obtained using a predefined step $\Delta$ in such a way that $|\textbf{m}_{k+1}-\textbf{m}_{k}| \propto \Delta$. Details of the generation of intermediate states are shown in Fig.~\ref{fig:IntermediateStates}. After obtaining energies $E_k$, the barrier height is obtained from $E_{AE} = \mathrm{max}\{E_2, E_3, ..., E_N\} - E_{init}$, where $E_N=E_{final}$. Finally, we proceed in a similar way to the Monte Carlo Metropolis algorithm for a classical spin system. The complete procedure is shown below:

1. Select a random atom $j$ with the initial direction of the magnetic moment  $\textbf{m}_{j,init}$. \\
2. Choose randomly a new trial direction $\textbf{m}_{j,final}$.\\ 
3. Generate intermediate magnetic states $\textbf{m}_{j,2},\textbf{m}_{j,3},...,\textbf{m}_{j,N-1}$ with the predefined step $\Delta$ on the arc connecting the initial $\textbf{m}_{j,init}=\textbf{m}_{j,1}$ and the final $\textbf{m}_{j,final}=\textbf{m}_{j,N}$ magnetization direction (see Fig.~\ref{fig:IntermediateStates}). \\
4. Calculate energies $E_{j,k}=f(\textbf{m}_{j,k})$ and the activation energy $E_{AE} = \mathrm{max}\{E_{j,2}, ..., E_{j,N}=E_{j,final}\} - E_{j,init}$\\
5. The new trial direction $\textbf{m}_{j,final}$ is accepted with the probability $P=$exp$(\frac{-E_{AE}}{k_BT})$, that is: \\
a. Draw a random number $r$ uniformly distributed between 0 and 1. \\
b. If $r$ $\leq$ exp$(\frac{-E_{AE}}{k_BT})$, accept the new state, otherwise reject it. \\
6. Repeat the procedure. \\

One complete iteration of sMC consists of $N_C$ such trial moves (one move corresponds to the whole procedure 1-5 presented above), where $N_C$ is the number of atoms in the system. In sMC approach we use $4\cdot 10^{7}$ equilibration and averaging iterations. 

Here for generation of the trial direction $\textbf{m}_{j,final}$ we choose randomly one of the two different trial moves : Gaussian or random (c.f. Fig.~3.b-c and 4.b in Ref.~\onlinecite{Evans:2014_JPhysCM} ). The Gaussian trial move creates $\textbf{m}_{j,final}$ by shifting the initial direction $\textbf{m}_{j,init}$ on the unit sphere according to the formula :

\begin{equation}
\label{eq:LLG}
\textbf{m}_{j,final}=\frac{ \textbf{m}_{j,init} + \sigma_g \bm{\Gamma} } {|\textbf{m}_{j,init} + \sigma_g \bm{\Gamma}|} ,
\end{equation}

where $\bm{\Gamma}$ is a 3D random vector characterized by the Gaussian distribution with a mean of zero and a standard deviation of 1. The width of the cone $\sigma_g$  takes the following form  $\sigma_g=\frac{2}{5}(\frac{k_BT}{\mu_B})^{1/5}$ \cite{Evans:2014_JPhysCM}, which aims to obtain a move acceptance rate of around 50~$\%$ . The random trial move creates a random point on the unit sphere $\textbf{m}_{j,final}= {\bm{\Gamma}}_U / |{\bm{\Gamma}}_U|$, where ${\bm{\Gamma}}_U$ is a 3D random vector with each components uniformly distributed between -1 and 1. In order to correctly (densely enough) arrange intermediate states (see Fig.~\ref{fig:IntermediateStates}) all trial states meeting the following condition $\textbf{m}_{j,init} \cdot \textbf{m}_{j,final} \leq -0.97$ are rejected.

We can see that the implementation of the sMC algorithm is a little more difficult than the standard MC one. Also the execution time for one sMC iteration is longer than in MC case, especially for small values of $\Delta$. However, the advantage of using the new method is that MC provides proper Boltzman distribution of states in relation to the global energy minimum, whereas sMC (in the short run) provides Boltzman distribution of states relative to the local energy minimum. This guarantees the correct dynamics of sMC, as energy barriers much higher than $k_BT$ prevent from too fast global thermalisation of spins.


It is important to note, that the sMC approach fulfills two general conditions for validity of Monte Carlo algorithms : ergodicity and the condition of detailed balance. Ergodicity expresses the requirement that all possible configurations of the system can be reached from any other configuration in a finite number of Markov steps. The detailed balance condition requires that, in the long run, the system approaches the correct thermal equilibrium. In sMC, for any two configurations $B$ and $C$ we have

\begin{equation}
\label{eq:balance}
\begin{split}
\frac{W(B \rightarrow C)}{W(C \rightarrow B)}=\frac{e^{-(E_{\mathrm{max}}-E_B)/(k_BT)}}{e^{-(E_{\mathrm{max}}-E_C)/(k_BT)}}= \\
=e^{-(E_C-E_B)/(k_BT)}=\frac{p(C)}{p(B)}
\end{split}
\end{equation}

where $W(B \rightarrow C)$ describes the probability to go from state $B$ to state $C$ in one move of the Markov process. $p(B)=e^{-E_B/(k_BT)}/Z$ is the Boltzmann probability distribution, in which $E_B$ is the energy of configuration $B$ and $Z$ is the partition function. $E_{\mathrm{max}}$ corresponds to the energy of the barrier (see Fig.~\ref{fig:sMC_schematic} with $B=\textbf{m}_{init}$ and $C=\textbf{m}_{final}$). Eq.~\ref{eq:balance} is also true in the case with no barrier, that is, when $E_B=E_{\mathrm{max}}$ or $E_C=E_{\mathrm{max}}$ (then sMC moves coincide with MC ones).


One should note, that the sMC approach presented here for spin systems can be generalized in a straightforward way to simulate other processes, for example diffusion, nucleation, surface adsorption, crystal growth processes and dynamics of classical atoms. Then we just generate intermediate states along the line connecting the initial and final (trial) positions. Moreover, to simulate the crystal growth, our model only requires knowledge of the Hamiltonian without having to assume a specific discrete structure of the crystal lattice.

\begin{figure}[htp]
    \centering
    \includegraphics[width=4cm]{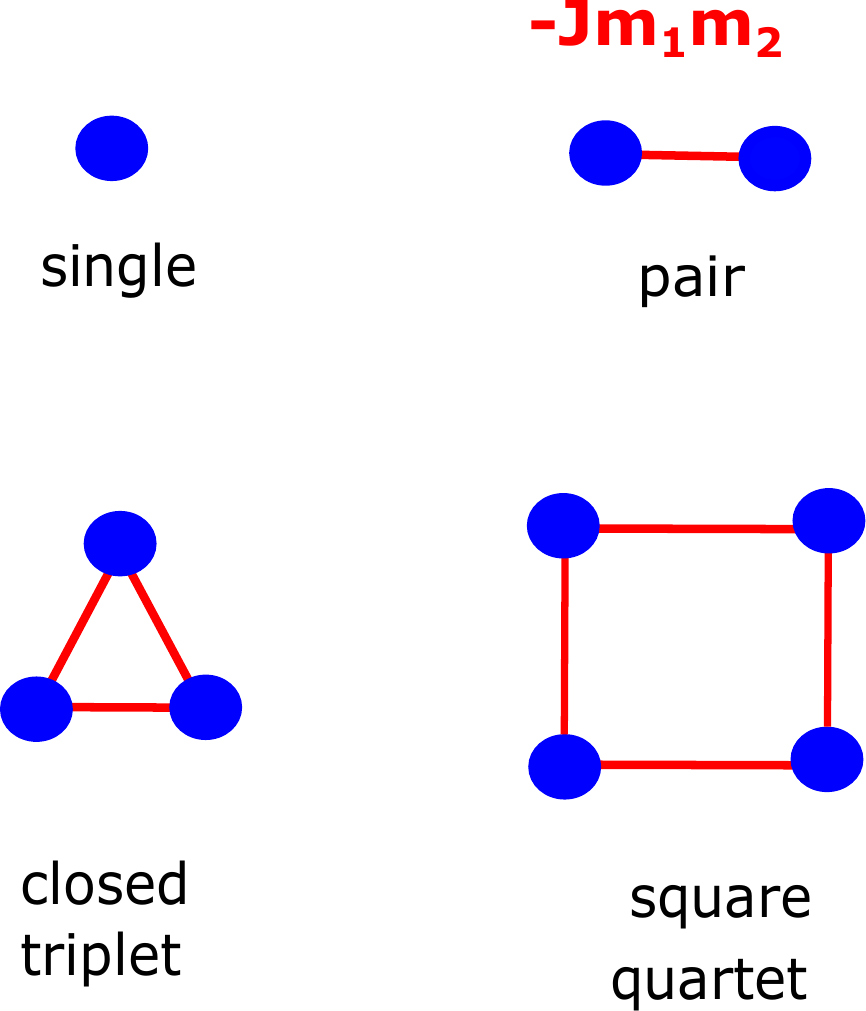}
    \caption{Investigated magnetic clusters composed of up to four magnetic ions (blue circles) coupled by a ferromagnetic exchange interaction (red lines).}
    \label{fig:clusters}
\end{figure}

To test the correctness of sMC, we compare its results with those obtained by stochastic Landau-Lifshitz-Gilbert (sLLG) equation \cite{Evans:2014_JPhysCM, Evans:2015_PRB, Edathumkandy:2022_JMMM}. Using the sLLG approach, it was possible to simulate the dynamic properties of magnetic materials \cite{Evans:2015_PRB} including ferromagnetic hysteresis loops \cite{Evans:2014_JPhysCM}, time evolution of a system of spins \cite{Evans:2014_JPhysCM}, spin waves \cite{Bender:2017_PRL}, domain wall motion \cite{Schlickeiser:2014_PRL, Hinzke:2011_PRL} and magnetization switching processes \cite{Ellis:2015_APL, Scholz:2001_JMMM}. The sLLG equation, applied at the atomistic level, reads

\begin{equation}
\label{eq:LLG}
\frac{\partial{\textbf{m}_i}}{\partial t}=- \frac{\gamma}{1+\alpha_G^2} [\textbf{m}_i\times\textbf{H}_{eff}^i+\alpha_G\textbf{m}_i\times(\textbf{m}_i\times\textbf{H}_{eff}^i)],
\end{equation}

Here $\gamma$ is the gyromagnetic ratio, $\alpha_G$ is the damping parameter. The symbol $\textbf{m}_i$ describes normalized ($|\textbf{m}_i|=1$) direction of the local magnetic moment of atom $i$ and  $\textbf{H}_{eff}^i$ corresponds to effective magnetic field acting on this spin.  $\textbf{H}_{eff}^i$ is calculated from the first derivative of the total spin Hamiltonian $\mathcal{H}$ and incorporates also the instantaneous thermal magnetic field $\textbf{H}_{Th}^i$, according to the following formula

\begin{equation}
\textbf{H}^i_{eff}=-\frac{1}{\mu_s}\frac{\partial \mathcal{H}}{\partial \textbf{m}_i}+\textbf{H}^i_{Th}
\end{equation}  

with

\begin{equation}
\label{eq:HThermal}
\textbf{H}_{Th}^i= \mathbf{\Gamma} (t) \sqrt{\frac{2 \alpha_G k_B T}{\gamma \mu_S \Delta t}}
\end{equation}

The instantaneous magnetic field $\textbf{H}_{Th}^i$ represents thermal effects (fluctuations) in the system, and allows to simulate the magnetization at a finite temperature $T$. $\mathbf{\Gamma}(t)$ is a random vector in 3D space characterized by the Gaussian distribution with a mean of zero and a standard deviation of 1. $\mu_S=g\mu_BS$ is an actual value of atomic moment with spin $S=2$, electron g-factor $g=2$ and $\Delta t = 5 \cdot 10^{-5}$~ns is the time step of simulation. Additionally, in sLLG method we use $\alpha_G = 1$ and $2 \cdot 10^{10}$ equilibration and averaging iterations.

\section{Comparison with stochastic LLG approach}


Having outlined the details of the step Monte Carlo method we now proceed to a comparative study of the magnetization dynamics obtained within the frame of the sMC and standard sLLG approach. We analyze a simple model consisting of few interacting magnetic atoms possessing the uniaxial anisotropy along the $z$ axis. The investigated small magnetic clusters are shown in Fig.~\ref{fig:clusters}. The Hamiltonian of the system reads

\begin{align}
\label{eq:H_toy}
\mathcal{H}=-k_z \sum_{i}{{m}_{iz}^2} - \mu_S\textbf{H}\sum_{i}{\textbf{m}_i} - J\sum_{i>j}{\textbf{m}_i}{\textbf{m}_j}
\end{align}

with appropriate terms relating to uniaxial anisotropy energy, Zeeman energy and the exchange coupling between atoms. Here $k_z=0.4$~meV is the anisotropy constant, $J=2$~meV describes the exchange parameter. The symbol $\textbf{H}$ corresponds to  the external magnetic field.

\begin{figure}[htp]
    \centering
    \includegraphics[width=8.6cm]{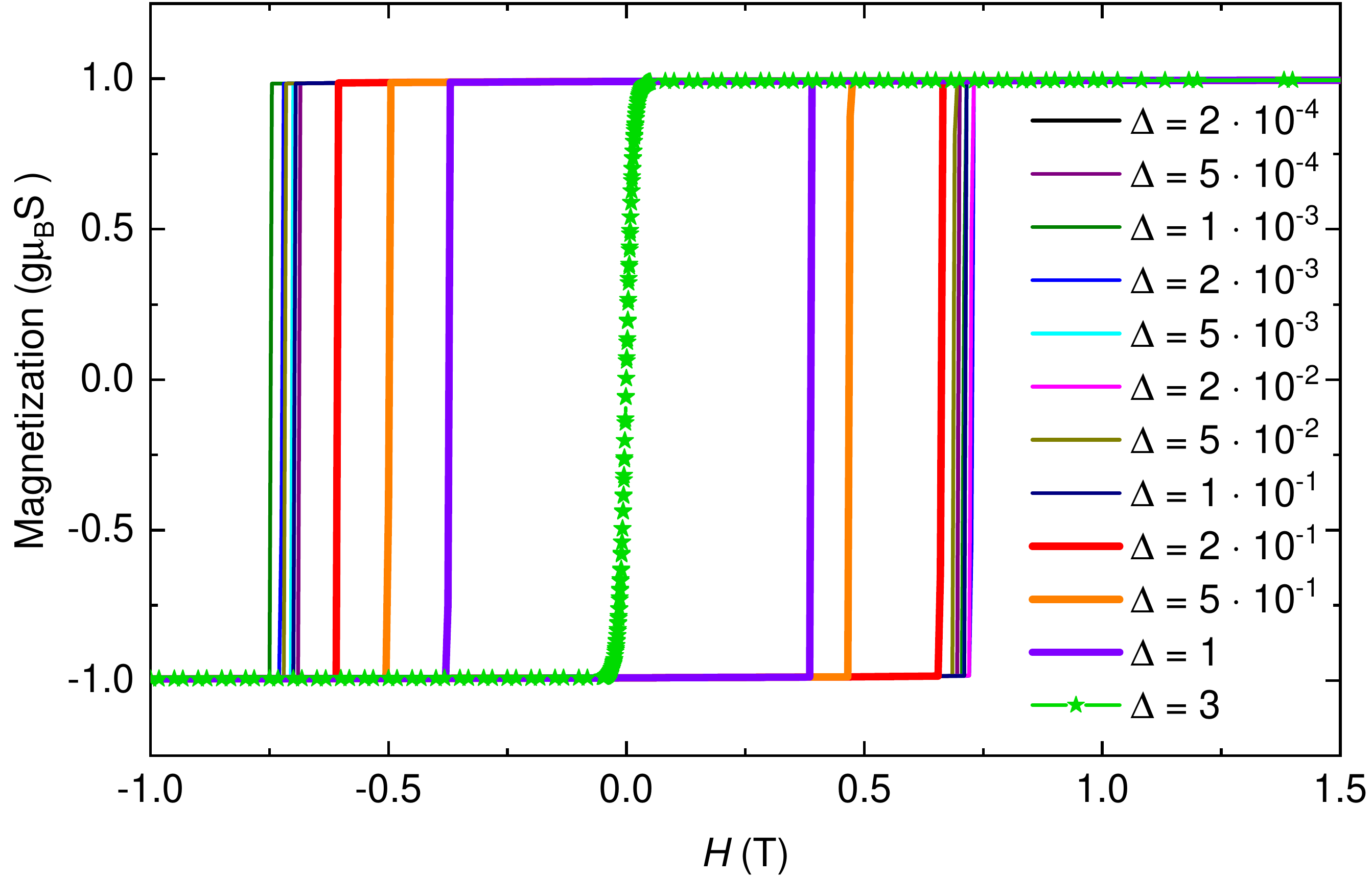}
    \caption{The magnetization $M$ of single ion, computed as a function of the external magnetic field $\textbf{H}$ applied along the magnetic easy axis at $T=0.04$~K. The simulations are performed using the sMC method with different steps $\Delta$. The standard MC approach is represented by $\Delta=3$. }
    \label{fig:MH_single}
\end{figure}

\begin{figure}[htp]
    \centering
    \includegraphics[width=8.6cm]{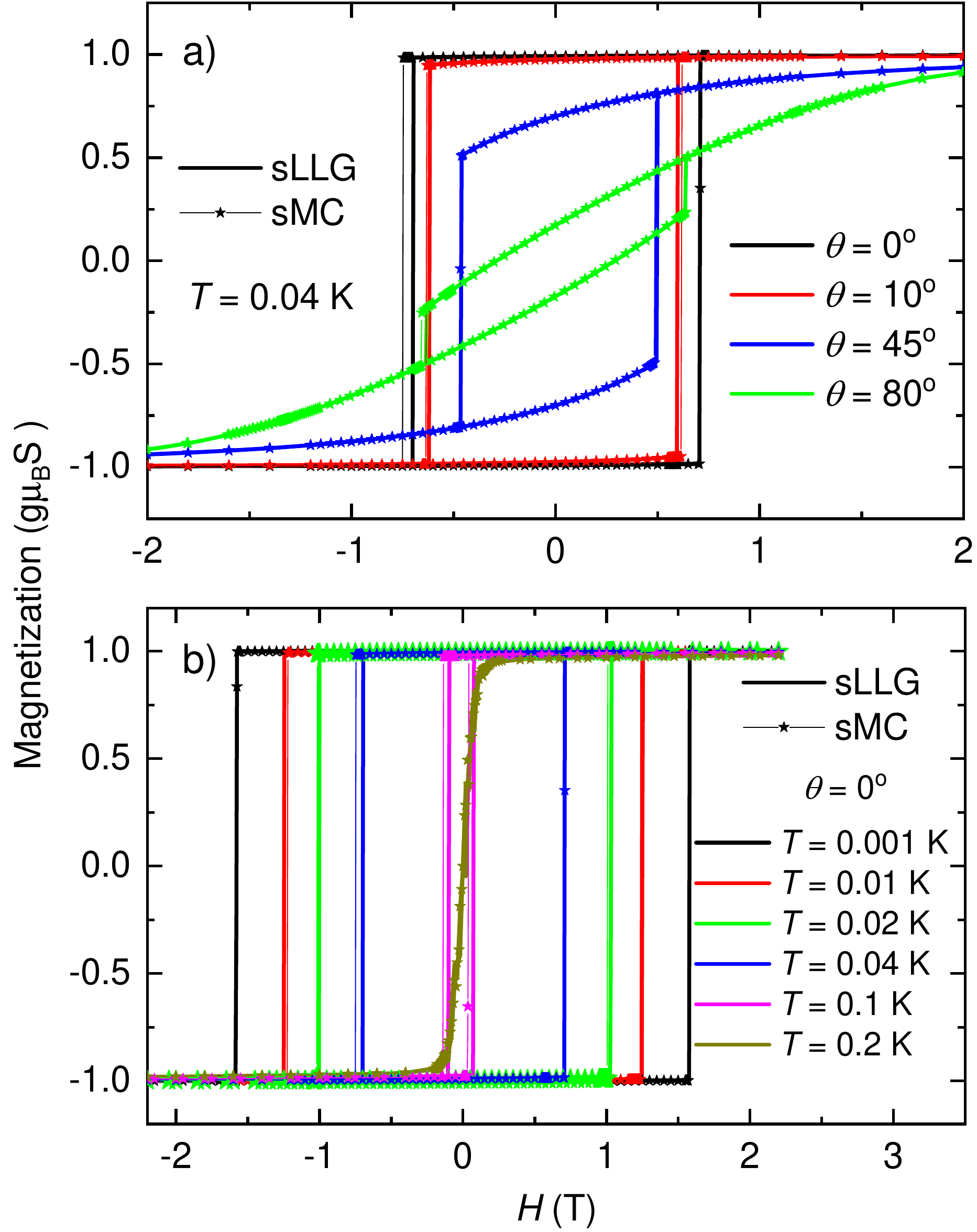}
    \caption{The comparison of numerical results between sMC ($\Delta = 10^{-3}$) and sLLG methods. a) The magnetization $M$ of a single ion with the magnetic field $\textbf{H}$ applied at different angles $\theta$ from the easy axis $\textbf{z}$ at temperature $T=0.04$~K. b) The magnetization $M$ of a single ion at different temperatures $T$ with the magnetic field $\textbf{H}$ applied along the easy axis $\textbf{H} || \textbf{z}$.}
    \label{fig:LM_comp_1ion}
\end{figure}

In Fig.~\ref{fig:MH_single} we present the magnetization $M$ of a single ion, computed as a function of the external magnetic field $\textbf{H}$ applied along the magnetic easy axis, namely $\textbf{H}$ $||$ $\textbf{z}$. The simulations are obtained using the sMC method with different steps $\Delta$. The standard MC approach is represented by $\Delta=3$. Only in this case we observe the rapid convergence to equilibrium state with no hysteresis loop seen in $M-H$ curve. By decreasing $\Delta$ critical dynamic behaviors far from equilibrium emerge. Now the magnetization has to overcome the energy barrier for switching, what creates the hysteresis loop. By decreasing $\Delta$ this barrier height is probed with higher resolution leading to the increase of the coercive field $H_C$. However, for $\Delta \leq 10^{-1}$ we observe saturation of $H_C$ and a clear presence of thermal fluctuations.

\begin{figure}[htp]
    \centering
    \includegraphics[width=8.6cm]{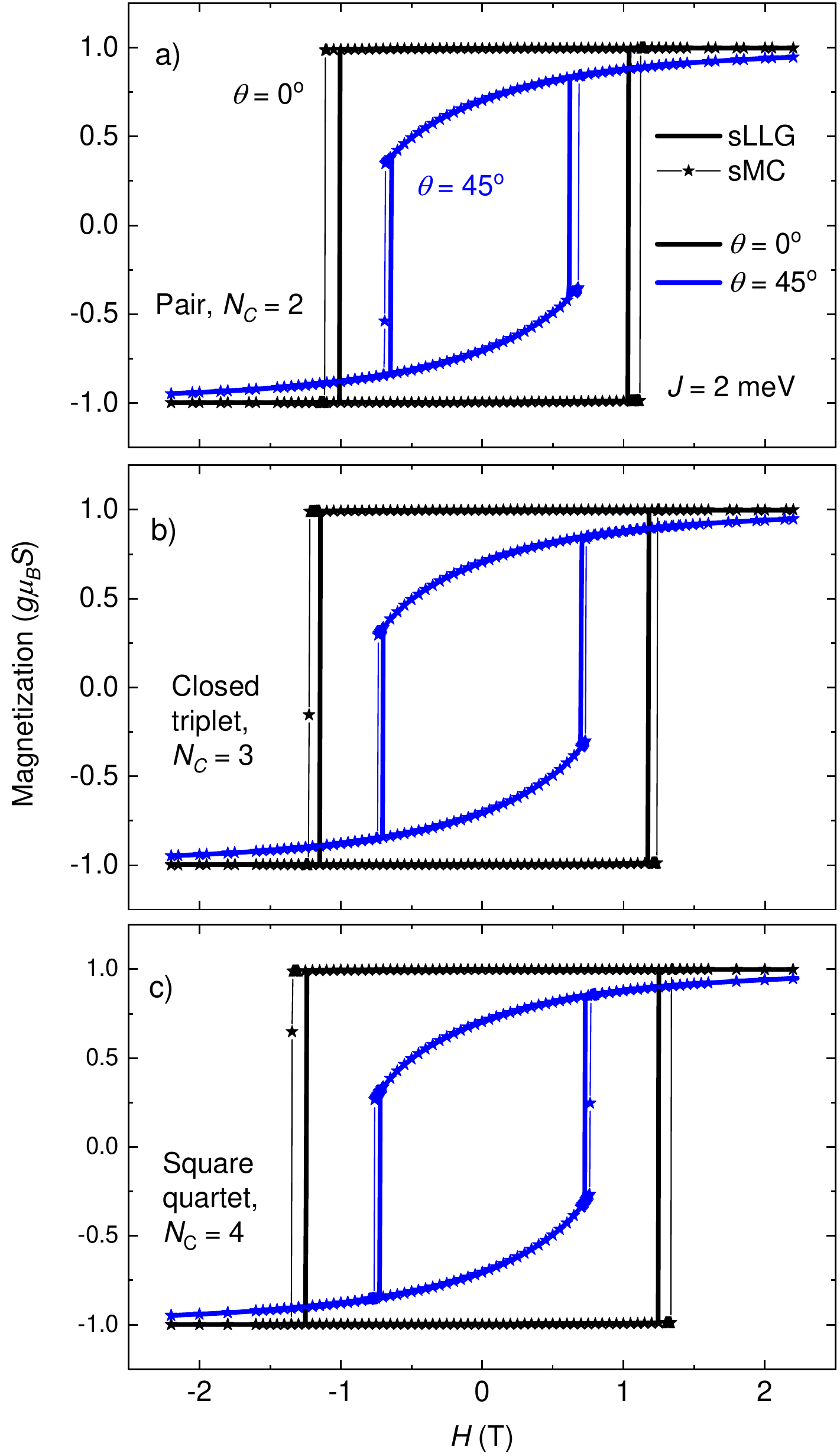}
    \caption{The comparison of numerical results between sMC ($\Delta = 10^{-3}$) and sLLG methods for different clusters with size $N_C$. The magnetization $M$ of a pair (a), closed triplet (b) and square quartet (c) at the temperature $T=0.04$~K. The magnetic field $\textbf{H}$ is applied at two different angles $\theta$ from the easy axis $\textbf{z}$, namely $\theta=0$ and $\theta=45^o$.}
    \label{fig:LM_comp_clusters}
\end{figure}

In Fig.~\ref{fig:LM_comp_1ion} the comparison of numerical results obtained using the sMC ($\Delta = 10^{-3}$) and sLLG method is displayed. First, in Fig.~\ref{fig:LM_comp_1ion} a. we plot the magnetization $M$ of a single ion with the magnetic field $\textbf{H}$ applied at different angles from the easy axis $\textbf{z}$ at the temperature $T=0.04$~K. As we see both approaches give very similar quantitative results. Small discrepancies are due to the presence of thermal fluctuations. These discrepancies decrease with lowering the temperature as shown in Fig.~\ref{fig:LM_comp_1ion} b. An additional source of the observed discrepancy between the results of sLLG and sMC is the fact, that the sLLG equation consists of two processes: precessional motion of the magnetization along the effective magnetic field and the damping term parametrized by $\alpha_G$. In the MC method only the last term, related to the relaxation of the system to equilibrium, is present.

Finally, magnetizations for larger clusters are presented in Fig.~\ref{fig:LM_comp_clusters}. At finite temperature, the coercive field increases slightly with the size of the cluster $N_C$, as expected. Again, the consistency between the sLLG and sMC results is very well, indicating a correct assumptions and implementation of the sMC algorithm.

Interestingly, the coercive field obtained from sLLG is slightly smaller than the one obtained from the sMC approach. This is especially noticeable for the magnetic filed applied along the easy axis $\textbf{H} || \textbf{z}$ ($\theta=0$). In our opinion, it is connected with the fact that in sMC alogirthm the Landau-Lifshitz (LL) term is absent. This LL term producing precessional motion of magnetization along the effective magnetic field seems to be important in such switching processes.

\section*{Conclusions}
	
Here, we present a modification of MC algorithm, named step Monte Carlo (sMC), that allows to simulate dynamic properties of various classical systems. In sMC the probability of accepting the final state depends on activation energy (barrier height), not on the relative energy between the final and initial state. This barrier height is probed by generating intermediate states  with a predefined step $\Delta$. To calculate the activation energy, our model only requires knowledge of the Hamiltonian without having to introduce additional input parameters such as transition rates etc. Importantly, the sMC method correctly takes into account the presence of various local barriers. As a result, the appropriate dynamics of the tested system is simulated. To test the correctness of sMC, we simulate dynamic magnetic properties for interacting spin system and compare its results with those obtained by stochastic Landau-Lifshitz-Gilbert (sLLG) approach. The obtained excellent correspondence of both methods indicates the correctness of the proposed approach. In our opinion, sMC method can be applied to simulate other processes, for example dynamics of classical atoms and complex fluids, diffusion, and crystal growth processes.

\section*{Acknowledgments}
I would like to thank M. Załuska-Kotur, C. Sliwa, C. Simserides and M. Sawicki for proofreading of the manuscript and valuable suggestions. The work is supported by the National Science Centre (Poland) through project OPUS (2018/31/B/ST3/03438).The calculations were made with the support of the Interdisciplinary Center for Mathematical and Computational Modeling of the University of Warsaw (ICM UW) under the computational grant no GB77-6.



\end{document}